\documentclass{article}
\usepackage{kr}

\usepackage{times}
\usepackage{soul}
\usepackage{url}
\usepackage[hidelinks]{hyperref}
\usepackage[utf8]{inputenc}
\usepackage[small]{caption}
\usepackage{graphicx}
\usepackage{complexity}

\usepackage{amsmath}
\usepackage{amsthm}
\usepackage{amssymb}
\usepackage{booktabs}
\usepackage{algorithm}
\usepackage{algorithmic}
\usepackage{todonotes}
\usepackage{braket}
\usepackage{mathtools}
\usepackage{xspace}

\newcommand{\var}{var}

\newcommand{\calA}{\mathcal{A}}

\newcommand{\calT}{\mathcal{T}}

\newcommand{\bbK}{\mathbb{K}}
\newcommand{\bbN}{\mathbb{N}}

\newcommand{\bbC}{\mathbb{C}}

\newtheorem{example}{Example}
\newtheorem{theorem}{Theorem}
\newtheorem{lemma}{Lemma}
\theoremstyle{definition}
\newtheorem{definition}{Definition}

\title{From Tensor Networks to Tractable Circuits, and back}

\author{%
Arend-Jan Quist$^1$\and
Marc Farreras Bartra$^{1,2}$\and
Alexis de Colnet$^{1}$\and
John van de Wetering$^2$\and\\
Alfons Laarman$^1$\\
\affiliations
$^1$Leiden University, Leiden, The Netherlands\\
$^2$University of Amsterdam, Amsterdam, The Netherlands\\
\emails
\{a.quist,a.e.h.de.colnet,a.w.laarman\}@liacs.leidenuniv.nl,
j.m.m.vandewetering@uva.nl
}

\begin{document}

\maketitle

\begin{abstract}
    Tensor networks and circuits are widely used data structures to represent pseudo-Boolean functions. These two formalisms have been studied primarily in separate communities, and this paper aims to establish equivalences between them. We show that some classes of tensor networks that are appealing in practice correspond to classes of circuits with specific properties that have been studied in knowledge compilation as \emph{tractable circuits}. In particular, we prove that matrix product states (tensor trains) coincide with nondeterministic edge-valued decision diagrams and that tree tensor networks exactly correspond to structured-decomposable circuits. These correspondences enable direct transfer of structural and algorithmic results; for example, canonicity and tractability guarantees known for circuits yield analogous guarantees for the associated tensor networks, and vice versa.
\end{abstract}

\section{Introduction}

Tensor networks and tractable circuits have emerged as two powerful systems for representing high-dimensional pseudo-Boolean functions.
Tensor networks, and in particular tensor trains (TTs or matrix product states)~\cite{Schollwock_MPS_DMRG,Oseledets_Tensor_Train} and tree tensor networks (TTNs)~\cite{Vidal_Tree_TN}, are mainly used in quantum physics, linear algebra, and increasingly in machine learning~\cite{martires2025quantum,LoconteMGPCQVV25,loconte2025sum}.
Tractable circuits, in various normal forms, such as decision diagrams (DDs)~\cite{Bryant_ROBDD} and DNNF~\cite{darwiche2001decomposable}, found extensive applications in formal verification, probabilistic analysis, and automated reasoning.
Despite their conceptual similarities, these formalisms have largely been developed separately by different communities.

In this paper, we establish key correspondences between the two formalisms. 
We show that the DD-equivalent of a TT is a non-deterministic (nd) extension of the well-known edge-valued DD (EVDD)~\cite{Lai_Integer_Linear_EVDD,Miller_DD_Qsim}.
The TTN, on the other hand, has a circuit equivalent in structured DNNF~\cite{pipatsrisawat2008new}, extended with edge weights. 
Based on these two correspondences, we then study weak and strong (decision) determinism, yielding syntactic or semantic properties that define the corresponding circuits and tensors. 
\autoref{tab:TN-DD_relations} summarizes our results.
These equivalences are constructive and allow bi-directional transformations with polynomial (often linear) overhead.

\begin{table}[b!]
	\renewcommand{\arraystretch}{1.2}
    \centering\footnotesize
    \setlength{\tabcolsep}{4pt}
    \begin{tabular}{p{11.2em}cp{11em}}
        \normalsize\textbf{Tensor network type} && \normalsize\textbf{Decision diagram type}\\
        \hline
        Tensor Train (TT) & $\Leftrightarrow$ & Non-Deterministic EVDD\\
        Deterministic TT %
        	& $\Leftrightarrow$ & EVDD\\
        Tree Tensor Network (TTN) & $\Leftrightarrow$ & EV-SDNNF \\
        Decision TTN & $\Leftrightarrow$ & Decision EV-SDNNF \\
        Deterministic TTN & $\Leftrightarrow$ & Deterministic EV-SDNNF\\
    \end{tabular}
    \caption{Tensor network and decision diagram equivalences.}
    \label{tab:TN-DD_relations}
\end{table}

These two main correspondences (TT $\equiv $ nd-EVDD and TTN $\equiv$ EV-SDNNF) enable cross-fertilization between two disciplines that have hitherto been largely separated.
For instance, the optimal variable order in DDs has been extensively studied and is known to be \NP-complete. Similarly, on the tensor network side, the size of a TT has been linked to the so-called Schmidt rank, and optimizing the order is also known to be hard.
Canonicity is inherent in the definition of (deterministic) DDs, but much harder to obtain for TTs, where one has to resort to singular value decomposition.
The tractability of operations has been extensively studied for circuits~\cite{DarwicheM02}, while different operations were considered for tensor networks~\cite{vinkhuijzen2024knowledge} and little is known about their composability~\cite{VergariCLTB21}. 
Finally, on the tensor network side, approximation by truncation is common practice, but it is not known yet what this means on the DD side, nor how it relates to approximate algorithms with rigorous theoretical guarantees for the same purpose~\cite{ArenasCJR21,meel2025fpras}. 
Our equivalences establish how these results relate to each other. 

In sum, this work contributes to a unified view of symbolic and algebraic representations for pseudo-Boolean functions, including settings with real, negative, or complex weights.
The provided equivalences can be used to transfer techniques, lower bounds, and algorithms between two well-developed but mostly separate fields. 
Additional opportunities for cross-fertilization between the tensor network community and the decision diagram community are discussed in more detail in Section~\ref{sec:discussion}.

\section{Preliminaries}

Semi-rings are written $\mathbb{K} = (S,+,\times,0,1)$, with $S$ the underlying set of elements. A Boolean variable is a variable that can take only two values: $0$ or $1$ (\emph{false} or \emph{true}).  A (truth) assignment to a set of Boolean variables $X$ is a mapping $\alpha \colon X \rightarrow \{0,1\}$. 
The set of assignments to $X$ is denoted by $\{0,1\}^X$. We use the notation $\{0,1\}^n$, $n \in \mathbb{N}$, to refer to the set of assignments to $\{x_1,\dots,x_n\}$.
A pseudo-Boolean function from $X$ to $\mathbb{K}$ is a mapping $f \colon \{0,1\}^X \rightarrow \mathbb{K}$. %

\subsection{Tractable Circuits}

We introduce classes of circuits to represent pseudo-Boolean functions. These circuits are sometimes referred to as \emph{tractable circuits} because they allow efficient poly-time procedures for several problems that are otherwise hard~\cite{DarwicheM02,VergariCLTB21}

\paragraph*{Pseudo-Boolean circuits.} A $(+,\times)$-circuit over $\mathbb{K} = (S,+,\times,0,1)$ is a directed acyclic graph (DAG) without multi-edges whose internal nodes, or \emph{gates}, are each labeled with the symbol $+$ and with the symbol $\times$ and whose sources, or inputs, are each labeled with an indicator variable $\mathbf{1}[x = 0]$ or $\mathbf{1}[x = 1]$ for $x$ a Boolean variable. Nodes with an outgoing edge to a gate $g$ are called the inputs or children~of~$g$. Gates have a finite but unbounded number of inputs. The set of variables of a gate $g$, denoted by $var(g)$ is the set of all $x$'s for which there is a circuit input $\mathbf{1}[x = b]$, $b \in \{0,1\}$, among the descendants of $g$. Each incoming edge to a $+$ gate is labeled with a value in $S$, called its weight. We write $g = w_1\cdot g_1 + \dots + w_k \cdot g_k$ to denote that $g$ is a $+$ gate with children $g_1,\dots,g_k$ and that the incoming edge from $g_i$ carries weight $w_i$. We write $g = g_1 \times \dots \times g_k$ to denote that $g$ is a $\times$ gate with children $g_1,\dots,g_k$. Each gate corresponds to a pseudo-Boolean function over $var(g)$ with a value in $S$. The value computed by $g$ on a (Boolean) assignment $\alpha$ to $var(g)$ is denoted by $g(\alpha)$. For $\times$ gates we have that $g(\alpha) = g_1(\alpha_1) \times \dots \times g_k(\alpha_k)$ and for $+$ gates we have that $g(\alpha) = w_1 \times g_1(\alpha_1) + \dots + w_k\times g_k(\alpha_k)$, where $\alpha_i$ is the restriction of $\alpha$ to $var(g_i)$. 
When $\mathbb{K}$ is the Boolean semi-ring $(\{0,1\},\lor,\land,0,1)$, the $(+,\times)$ circuits coincide with the \emph{circuits in negation normal form} (NNF). When $\mathbb{K}$ is the real field $(\mathbb{R},+,\times,0,1)$ they are the \emph{arithmetic circuits}. When $\mathbb{K}$ is the semi-ring of the non-negative reals $(\mathbb{R}_+,+,\times,0,1)$ they are the \emph{monotone arithmetic circuits}.

\paragraph{Determinism.} A $(+,\times)$ circuit is \emph{deterministic} when, for every $+$ node $w_1\cdot g_{1} + \cdots + w_k \cdot g_{k}$ we have that $w_i\times g_i \times w_j \times g_j$ is the $0$ function for every $i \neq j$. So either $w_i$ or $w_j$ is the $0$ weight, or $g_i \times g_j$ is the $0$ function. 

\paragraph{Decomposability.} A $(+,\times)$ circuit is \emph{decomposable} when, for every $\times$ gate $g = g_1 \times \dots \times g_k$, it holds that $\var(g_i) \cap \var(g_j) = \emptyset$ for every $1 \leq i < j \leq k$. In a sense, in a decomposable circuit, the $\times$ splits the variables.

\begin{figure*}[t!]
    \centering
    \begin{minipage}{.4\textwidth}
        \begin{tikzpicture}[xscale=1.3,yscale=1.2]
            \node[circle,draw,inner sep=2pt] (+1) at (.5,1) {$+$};
            \node[circle,draw,inner sep=2pt] (+2) at (2.5,1) {$+$};
            
            \node[circle,draw,inner sep=2pt] (x1) at (0,0){$\times$};
            \node[circle,draw,inner sep=2pt] (x2) at (1,0){$\times$};
            \node[circle,draw,inner sep=2pt] (x3) at (2,0){$\times$};
            \node[circle,draw,inner sep=2pt] (x4) at (3,0){$\times$};

            \node[circle,draw,inner sep=2pt] (1+) at (0-.2-.2,-1) {$+$};
            \node[circle,draw,inner sep=2pt] (2+) at (1-.2-.2,-1) {$+$};
            \node[circle,draw,inner sep=2pt] (3+) at (2-.2,-1) {$+$};
            \node[circle,draw,inner sep=2pt] (4+) at (3-.2,-1) {$+$};
            \node[circle,draw,inner sep=2pt] (5+) at (4-.2,-1) {$+$};

            \node[rectangle, inner sep=2pt,font=\scriptsize] (varx1) at (1-.2-.2,-2) {$\mathbf{1}[x_1 = 1]$};
            \node[rectangle, inner sep=2pt,font=\scriptsize] (var-x1) at (0-.2-.2,-2) {$\mathbf{1}[x_1 = 0]$};
            \node[rectangle, inner sep=2pt,font=\scriptsize] (varx2) at (2.5-.2,-2) {$\mathbf{1}[x_2 = 1]$};
            \node[rectangle, inner sep=2pt,font=\scriptsize] (var-x2) at (3.5-.2,-2) {$\mathbf{1}[x_2 = 0]$};

            \draw[red] (+1) to node[midway,fill=white,inner sep=1pt,font=\footnotesize] {$1$} (x1);
            \draw[red] (+1) to node[midway,fill=white,inner sep=1pt,font=\footnotesize] {$1$} (x2);
            \draw[red] (+1) to node[midway,fill=white,inner sep=1pt,font=\footnotesize] {$-1$} (x3);
            \draw (+2) to node[midway,fill=white,inner sep=1pt,font=\footnotesize] {$1$} (x2);
            \draw (+2) to node[midway,fill=white,inner sep=1pt,font=\footnotesize] {$2$} (x4);
            \draw (+2) to node[midway,fill=white,inner sep=1pt,font=\footnotesize] {$0.5$} (x3);

            \draw (x1) -- (1+);
            \draw (x1) -- (3+);
            \draw (x2) -- (2+);
            \draw (x2) -- (4+);
            \draw (x3) -- (2+);
            \draw (x3) -- (5+);
            \draw (x4) -- (1+);
            \draw (x4) -- (4+);

            \draw (1+) to node[midway,fill=white,inner sep=1pt,font=\footnotesize] {$1$} (var-x1);
            \draw (2+) to node[midway,fill=white,inner sep=1pt,font=\footnotesize] {$3$} (varx1);
            \draw (3+) to node[midway,fill=white,inner sep=1pt,font=\footnotesize] {$2$} (varx2);
            \draw[blue] (4+) to node[midway,fill=white,inner sep=1pt,font=\footnotesize] {$1.5$} (varx2);
            \draw[blue] (4+) to node[midway,fill=white,inner sep=1pt,font=\footnotesize] {$1$} (var-x2);
            \draw (5+) to node[midway,fill=white,inner sep=1pt,font=\footnotesize] {$-1$} (var-x2);

        \end{tikzpicture}
    \end{minipage}
    \begin{minipage}{0.25\textwidth}
        \begin{tikzpicture}[level distance=1.5cm,
        level 2/.style={sibling distance=2cm}]
            \tikzstyle{every node}=[rectangle]
            \node [white] {}
                child {
                node {$v$} 
                child { node {$v_\ell$} edge from parent node[left,draw=none,font=\footnotesize] {$d(v_\ell)=2$} }
                child { node {$v_r$} edge from parent node[right,draw=none,font=\footnotesize] {$d(v_r)=3$}}
                edge from parent node[left,draw=none,font=\footnotesize] {$d(v)=2$}
            };
        \end{tikzpicture}
    \end{minipage}
    \begin{minipage}{0.25\textwidth}
        \begin{align*}
            \calA(v)&=\left(
            {\color{red}\begin{pmatrix}
                1&0&0\\
                0&1&-1
            \end{pmatrix}}
            \begin{pmatrix}
                0&2&0\\
                0&1&0.5
            \end{pmatrix}
            \right)
            \\
            \calA(v_\ell)&=\left(
            \begin{pmatrix}
                0\\
                1
            \end{pmatrix}
            \begin{pmatrix}
                3\\
                0
            \end{pmatrix}
            \right)\\
            \calA(v_r)&=\left(
            \begin{pmatrix}
                2\\
                0
            \end{pmatrix}
            {\color{blue}\begin{pmatrix}
                1.5\\
                1
            \end{pmatrix}}
            \begin{pmatrix}
                0\\
                -1
            \end{pmatrix}
            \right)
        \end{align*}
    \end{minipage}
    \caption{Example of a structured-decomposable circuit (left) and its corresponding tree tensor network (right). The matrices of the tensors correspond to adjacency matrices between the layers of the circuit.}
    \label{fig:TTN-SDD}
\end{figure*}
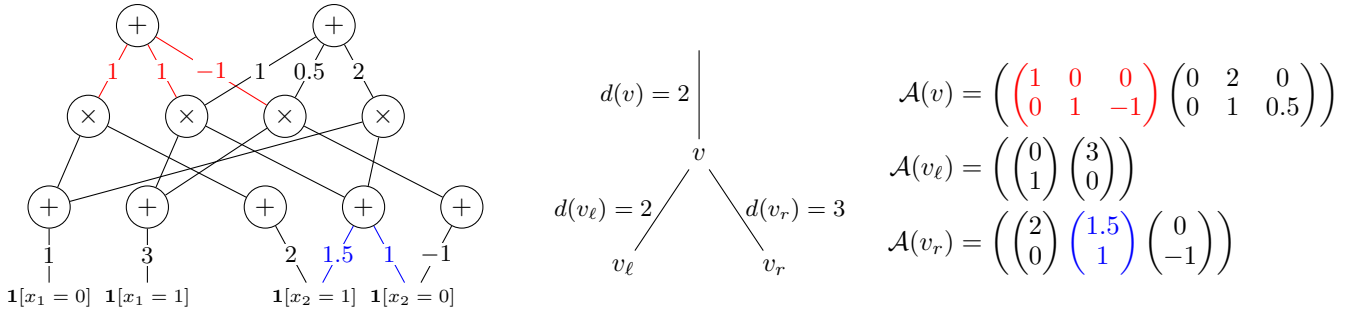

\paragraph{Structured-decomposability.}
In this paper, we consider decomposable circuits that can only split variables in a specific way. The legal splits are encoded in a \emph{vtree} (variable tree). Let $X$ be a set of Boolean variables. A \emph{vtree} over $X$ is a pair $(T,\lambda)$ with $T$ a rooted binary tree and $\lambda : leaves(T) \rightarrow X$ a bijection. For every $v \in V(T)$, we denote by $\var(v)$ the set of variables on the leaves of the subtree rooted at $v$. Formally, $\var(v) = \lambda(v)$ when $v \in leaves(T)$, and $\var(v) = \var(v_\ell) \cup \var(v_r)$ when $v$ is an internal node with children $v_\ell$ and $v_r$. A $(+,\times)$ circuit $C$ is \emph{structured by the vtree} $(T,\lambda)$ over $\var(C)$ when there is a mapping $\phi$ from $C$'s gates to $V(T)$ such that 
\begin{itemize}
\item[•] gates $g$ map to nodes $\phi(g)$ such that $var(g) \subseteq var(\phi(g))$
\item[•] each $\times$ gate $g$ has exactly two children $g_\ell$ and $g_r$, $\phi(g)$ is an internal node $v$ of $T$ and there exist $v'_\ell \in \{v_\ell\} \cup descendants(v_\ell)$ and  $v'_r \in \{v_r\} \cup descendants(v_r)$ such that $\phi(g_\ell) = v'_\ell$ and $\phi(g_r) = v'_r$.
\item[•] each $+$ gate $g = w_1 \cdot g_1 + \dots + w_k \cdot g_k$ is such that $\phi(g_i)$ is either $\phi(g)$ or a descendant of $\phi(g)$. 
\end{itemize}
A circuit structured by a vtree is \emph{structured-decomposable}. Structured-decomposability implies decomposability.

\subsection*{Edge Valued Decision Diagrams (EVDD)}

Edge-valued decision diagrams are popular representations for pseudo-Boolean functions to a semi-ring. We review some classes of edge-valued decision diagrams and recall that, in a sense, they are subsumed by classes of circuits. 

\paragraph*{Read-once EVDD.} A read-once edge-valued decision diagram over $\bbK$ is a rooted DAG $G$ with a single source and a single sink. Each internal node is labeled with a Boolean variable and has either one outgoing edge labeled $0$ or $1$, or two outgoing edges, one labeled $0$ and the other labeled $1$. The endpoint of a $0$-edge (resp. $1$-edge) is called the $0$-child (resp. $1$-child). The $0$-child and the $1$-child may be identical. Each edge $e$ is labeled with a value $w(e) \in \bbK$. $G$ is \emph{read-once} in the sense that no variable $x$ labels more than one node along the same source-to-sink path. A read-once EVDD is used to represent a function $f \colon \{0,1\}^X \rightarrow \bbK$ with $X$ the finite set of variables labeling its internal nodes. For every assignment $\alpha$ to $X$, we follow the unique path $p_\alpha$ in $G$ starting at the source of $G$ and following the $\alpha(x)$-edge whenever encountering a node labeled with $x$. Let $E(p_\alpha)$ be the set of edges in $p_\alpha$. If $p_\alpha$ reaches the sink then $f(\alpha) = \prod_{e \in E(p_\alpha)} w(e)$. Otherwise $f(\alpha) = 0$. The set of variables labeling internal nodes of $G$ is denoted by $\var(G)$.

An \emph{ordered} EVDD is a read-once EVDD where there is a total ordering $\prec$ on $\var(G)$ respecting the variable order along all paths, i.e., if $v$ is an ancestor of $u$ then $x_v \prec x_u$.

\paragraph*{Non-deterministic EVDD.} Non-deterministic read-once EVDD (nd-EVDD) is defined like read-once EVDD except that each node can have several $0$-edges and several $1$-edges. The definitions of read-once and ordered properties translate to non-deterministic EVDD in a natural way. However, we need to explain how to interpret the diagram as a pseudo-function $f$. For an assignment $\alpha$ to $X$ (with $X \supseteq \var(G)$) we now have a set $P_\alpha$ of paths associated with $\alpha$, namely, all paths starting from the source, ending at the sink, and that follow an $\alpha(x)$-edge whenever encountering a node labeled with $x$, then we have $f(\alpha) = \sum_{p \in P_\alpha}\prod_{e \in E(p)} w(e)$. If $P_\alpha = \emptyset$ and $f(\alpha) = 0$. In the paper, we mainly focus on non-deterministic ordered EVDD, so we will refer to them as EVDD when the context is clear.

\subsection{Tensor Networks}

In the following, $\mathbb{K} = (S,+,\times,0,1)$ is a semi-ring. We write $\times$ as $\cdot$ if the context is clear.

\paragraph*{Tensors.} A \emph{tensor} $A$ of order $r$ with dimensions $d_1 \times \cdots \times d_r$ over $\mathbb{K}$ is an element in $\mathbb{K}^{d_1 \times \cdots \times d_r}$. The components, or \emph{entries}, of $A$ are denoted by $A[i_1,\dots,i_r]$ where $i_k \in [d_k]$ for every $k \in [r]$. When a dimension is $2$ we sometimes take its entry index in $\{0,1\}$ rather than in $\{1,2\}$. Given $A$ and $B$ two tensors over $\mathbb{K}$ of the same dimension $d_1 \times \cdots \times d_r$, we denote by $A \odot B$ their element-wise product, that is, $A \odot B$ is a tensor over $\mathbb{K}$ of dimension $d_1 \times \cdots \times d_r$ such that $(A \odot B)[i_1,\dots,i_r] = A[i_1,\dots,i_r] \times B[i_1,\dots,i_r]$.

\emph{Tensor contraction}  generalizes matrix multiplication. Let $A$ and $B$ be two tensors over $\mathbb{K}$ with dimensions $d_1 \times \cdots \times d_r$ and $n_{1} \times \cdots \times n_{s}$ respectively. Suppose $d_k = n_l$ . The contraction of $k^\text{th}$ dimension of $A$ with the $l^\text{th}$ dimension of $B$ is the tensor over $\mathbb{K}$ denoted by $A \ast_{k,l} B$ of dimensions $d_1 \times \cdots \times d_{k-1} \times d_{k+1} \times \dots \times d_r \times n_{1} \times \dots \times n_{l-1} \times n_{l+1} \times \dots \times n_{s}$ whose entries are as follows, for $\vec{e}_1 \in [d_1] \times \cdots \times [d_{k-1}]$, $\vec{e}_2 \in [d_{k+1}] \times \dots \times [d_r]$, $\vec{e}_3 \in [n_{1}] \times \dots \times [n_{l-1}]$, $\vec{e}_4 \in [n_{l+1}] \times \dots \times [n_{s}]$,
$$
(A \ast_{k,l} B)[\vec{e}_1,\vec{e}_2,\vec{e}_3,\vec{e}_4] = \sum_{a \in [d_k]} A[\vec{e}_1,a,\vec{e}_2]\cdot B[\vec{e}_3,a,\vec{e}_4]
$$

A function $f \colon \{0,1\}^n \rightarrow \mathbb{K}$ can be represented by the order-$n$ tensor $A \in \mathbb{K}^{2 \times \cdots \times 2}$ with entries $A[\alpha] = f(\alpha)$ for $\alpha \in \{0,1\}^n$. We may use other tensor representations of $f$ equivalent under tensor isomorphisms. For instance, $f$ may be represented as a vector in $\mathbb{K}^{2^n}$. Furthermore, a collection of $d$ functions $(f^{(1)},\dots, f^{(d)}) \in (\{0,1\}^{n} \rightarrow \mathbb{K})^d$ can be jointly represented as a tensor over $A \in \mathbb{K}^{2 \times \cdots \times 2 \times d}$ with entries $A[\alpha ,i] = f^{(i)}(\alpha )$ where $\alpha \in \{0,1\}^n$ and $i \in [d]$.

\paragraph*{Tensor trains (TTs).} 
A \emph{tensor train} (TT), also called a matrix product state (MPS), is an ordered array $\calA$ of $n$ tensors, denoted as $\calA{(r)}$ with $r \in [n]$.
Each tensor $\calA{(r)}$ with $r\in[n]$ has components $\calA{(r)}[j_r,j_{r+1},i]$, with $i\in\{0,1\}$. The indices $j_r$ and $j_{r+1}$ are called virtual indices, with $j_r\in[\chi_r]$, $j_{r+1}\in[\chi_{r+1}]$, where $\chi_r$ is an integer for $r\in[n]$. We constrain $\chi_1=\chi_{n+1}=1$, and remove this dimension if the context is clear. 
So, tensors at the ends have order two, while others have order three. 
The maximum over $\chi_r$ is called the \textit{bond dimension} of the tensor train. The tensor train $\calA$ represents the tensor $\hat\calA$ over $\bbK^{2^n}$, seen as a function $f : \{0,1\}^n \mapsto \mathbb{K}$ with $f(\alpha)$ defined as
$$\hat\calA[\alpha] =(\calA(1)\ast_{2,1}\calA(2)\ast_{2,1}\dots\ast_{2,1}\calA(n-1)\ast_{2,1}\calA(n))[\alpha],$$
where the variable $x_r$ belongs to tensor $\calA(r)$.

\paragraph*{Tree tensor networks.}

A \emph{binary tree tensor network (TTN)} is a tuple $(\calA, T, d)$ where $T$ is a rooted binary tree directed from the root to the leaves, $\calA$ is a mapping from $V(T)$ to tensors over $\bbK$ and $d$ maps $V(T)$ to $\bbN$. If $v$ is a leaf of $T$ then $\calA(v)$ is a $2 \times d(v)$ tensor over $\bbK$. If $v$ is an internal node of $T$ with children $v_\ell$ and $v_r$ then $\calA(v)$ is a $d(v_\ell) \times d(v_r) \times d(v)$ tensor over $\bbK$. We define $\hat\calA(v)$ has $\hat\calA(v) = \calA(v)$ for leaves of $T$ and $\hat\calA(v) = \calA(v) \ast_{1,2} \hat\calA(v_\ell)  \ast_{2,2} \hat\calA(v_r)$ for internal nodes. One can check $\hat\calA(v)$ is a $2 \times \dots \times 2 \times d(v)$ tensor over $\bbK$ where the number of $2$ is the number of leaves below $v$ (including $v$) in $T$. It will be convenient to see $\hat\calA(v)$ as a $2^k \times d(v)$ tensor and thus $\hat\calA(v)[\cdot,i]$, for $i \in [d(v)]$ as a pseudo-Boolean function over $k$ variables. Therefore, a $n$-leaves binary TTN over $\bbK$ computes a sequence of $d(\text{root}(T))$ functions in $\{0,1\}^n \rightarrow \bbK$.
We will refer to binary TTNs simply as TTNs, since we only consider this form.

\begin{example}
Figure~\ref{fig:TTN-SDD} provides, on the right, an example of TTN with $T$ a binary tree with $3$ nodes: $v$, $v_\ell$, and $v_r$. One can check that $\hat\calA(v)$ is 
$$
\left(
            {\begin{pmatrix}
                0&2\\
                6&4.5
            \end{pmatrix}}
            \begin{pmatrix}
                2&3\\
                1.5&4.5
            \end{pmatrix}
            \right)
$$
So the TTN represents two pseudo-Boolean functions: the first maps $(x_1 = 0, x_2 = 0)$ to $0$, $(x_1 = 0, x_2 = 1)$ to $2$, etc. The second function maps $(x_1 =0,x_2 = 0)$ to $2$, $(x_1 =0,x_2 = 1)$ to $3$, etc.
\end{example}

\section{From Tensor Train to (Non-deterministic) EVDD, and back}

We explain the equivalence between TT and non-deterministic EVDD, and how to efficiently transform between them. Moreover, we show that a TT where all matrices in the tensor train have at most one non-zero entry per row corresponds to a deterministic EVDD. Figure~\ref{fig:nEVDD_example} shows an example of how TT and nd-EVDD relate to each other.

\tikzstyle{startstop} = [rectangle, rounded corners, 
minimum width=3cm, 
minimum height=1cm,
text centered, 
draw=black, 
fill=red!30]

\tikzstyle{io} = [trapezium, 
trapezium stretches=true, 
trapezium left angle=70, 
trapezium right angle=110, 
minimum width=3cm, 
minimum height=1cm,
text centered, 
text width=4cm,
draw=black, fill=blue!30]

\tikzstyle{process} = [rectangle, 
minimum width=3cm, 
minimum height=1cm, 
text centered, 
text width=4cm, 
draw=black, 
fill=orange!30]

\tikzstyle{decision} = [diamond, 
minimum width=3cm, 
minimum height=1cm, 
text centered, 
draw=black, 
fill=green!30]
\tikzstyle{arrow} = [thick,->,>=stealth]

\usetikzlibrary{patterns,shapes,arrows,fit,calc,positioning,automata}
\tikzset{unused/.style={pattern = north east lines}}
\tikzset{main/.style={draw,circle}} %
\tikzset{leaf/.style={draw,minimum width=1.2em,minimum height=1.2em}} %
\tikzset{e0/.style={draw,->,dotted,>=latex}}
\tikzset{e1/.style={draw,->,>=latex}}
\tikzset{XOR/.style={draw,circle,append after command={
        [shorten >=\pgflinewidth, shorten <=\pgflinewidth,]
        (\tikzlastnode.north) edge (\tikzlastnode.south)
        (\tikzlastnode.east) edge (\tikzlastnode.west)
        }
    }
}
\newcommand{\inlinecnot}
{\protect\tikz[baseline=.1ex]{
\protect\node[XOR,minimum size=1.3ex,inner sep=0cm] (A) at (0,0){};
\protect\fill (0,1.5ex) circle (1.5pt) coordinate (B);
\protect\draw [line width=.2mm] (A)--(B);}
\hspace{-.08cm}}

\newcommand{\leafsymb}{1}
\newcommand{\inlineleaf}
{\protect\tikz[baseline=-.6ex, every node/.style={scale=0.85}]{
\protect\node[leaf] (0) at (0,0){\normalfont{1}};
}}
\newcommand{\inlineedge}[2]
{\protect\tikz[baseline=-.5ex,every node/.style={scale=0.85}]{
\protect\node[inner sep=0mm] (A) at (0,0){};
\protect\node[main,minimum width=4.5mm,inner sep=0mm] (B) [right = .5cm of A] {\(#1\)};
\protect\draw[->,>=latex] (A) to node [above] {\(#2\)} (B);
}}
\newcommand{\inlineleafedge}[1]
{\protect\tikz[baseline=-.5ex,every node/.style={scale=0.85}]{
\protect\node[inner sep=0mm] (A) at (0,0){};
\protect\node[leaf] (B) [right = .6cm of A] {\leafsymb};
\protect\draw[->,>=latex] (A) to node [above,pos=0.4] {\(#1\)} (B);
}}
\renewcommand{\leafsymb}{\top}
\renewcommand{\inlineleaf}{\leafsymb}
\renewcommand{\inlineedge}[2]{\textnormal{\textsc{Edge}}(#1, #2)}
\renewcommand{\inlineleafedge}[1]{\textnormal{\textsc{Edge}}(\leafsymb, #1)}

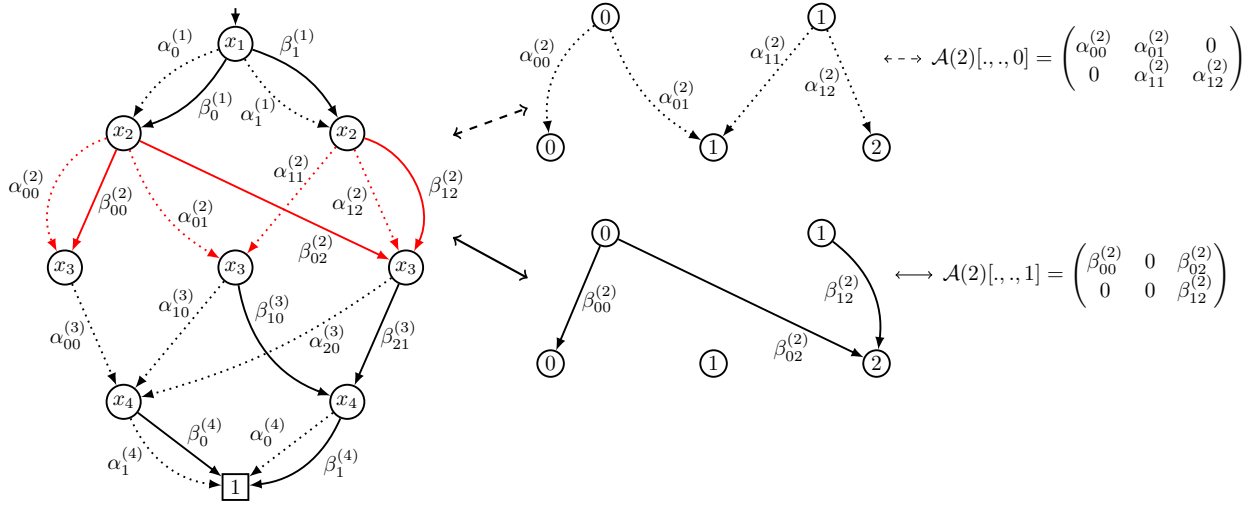
\begin{figure*}[t!]
    \centering
    \begin{tikzpicture}[scale=.9, every node/.style={scale=.9}]
    \node (left_fig) at (0.5,0) {\begin{tikzpicture}[auto, thick,node distance=3.cm,inner sep=1.5pt,scale=0.8]
    \newcommand\dist{.33cm}
    \newcommand\sep{.7cm}
    \newcommand\sepp{0.15cm}
    \newcommand\seppp{-.15cm}
    \node[] (0) [] {};
    \node[main] (x0) [node distance=.65cm,below of=0] {$x_1$};
    \node[] (1) [below = 3*\dist of x0] {};  
    \node[main] (x10)  [left = 4*\dist of 1] {$x_2$};
    \node[main] (x11)  [ right = 4*\dist of 1] {$x_2$};
    \node[main] (x21) [below = 5*\dist of 1] {$x_3$};
    \node[main] (x20) [left  = 6*\dist of x21] {$x_3$};
    \node[main] (x22) [right  = 6*\dist of x21] {$x_3$};
    \node[] (3) [below = 5*\dist of x21] {};
    \node[main] (x30) [left = 4*\dist of 3] {$x_4$};
    \node[main] (x31) [right = 4*\dist of 3] {$x_4$};
    \node[leaf] (l)   [below = 3*\dist of 3] {$1$};
    \draw[e1] (0) to (x0);
    \draw[e0,bend right=20] (x0) to node [above left=-0.05,pos=0.2] {\hspace{.5mm}$\alpha_{0}^{(1)}$} (x10);
    \draw[e1,bend left=20] (x0) to node [below right=-0.15,pos=0.4] {\hspace{.5mm}$\beta^{(1)}_{0}$} (x10);
    \draw[e1,bend left=20] (x0) to node [above right=-0.05,pos=0.2] {\hspace{.5mm}$\beta^{(1)}_{1}$} (x11);
    \draw[e0,bend right=20] (x0) to node [below left=-0.1,pos=0.45] {\hspace{.5mm}$\alpha_{1}^{(1)}$} (x11);
    
    \draw[e0,bend right=50, draw=red] (x10) to node [left=0.1,pos=0.5] {\hspace{.5mm}$\alpha^{(2)}_{00}$} (x20);
    \draw[e1, draw=red] (x10) to node [right=-0.05,pos=0.5] {\hspace{.5mm}$\beta^{(2)}_{00}$} (x20);
    \draw[e0,bend right=20, draw=red] (x10) to node [right=0.1,pos=0.5] {\hspace{.5mm}$\alpha^{(2)}_{01}$} (x21);
    \draw[e1, draw=red] (x10) to node [below=0.1,pos=0.7] {\hspace{.5mm}$\beta^{(2)}_{02}$} (x22);

    \draw[e1,bend left=50, draw=red] (x11) to node [right=0.1,pos=0.5] {\hspace{.5mm}$\beta^{(2)}_{12}$} (x22);
    \draw[e0, draw=red] (x11) to node [left=0.05,pos=0.5] {\hspace{.5mm}$\alpha^{(2)}_{12}$} (x22);
    \draw[e0, draw=red] (x11) to node [left=0.05,pos=0.2] {\hspace{.5mm}$\alpha^{(2)}_{11}$} (x21);

    \draw[e0] (x20) to node [left=0.05,pos=0.5] {\hspace{.5mm}$\alpha^{(3)}_{00}$} (x30);

    \draw[e0] (x21) to node [left=0.1,pos=0.2] {\hspace{.5mm}$\alpha^{(3)}_{10}$} (x30);
    \draw[e1,bend right=30] (x21) to node [right=+0.05,pos=0.15] {\hspace{.5mm}$\beta^{(3)}_{10}$} (x31);

    \draw[e0, bend left=10] (x22) to node [right=0.1,pos=0.4] {\hspace{.5mm}$\alpha^{(3)}_{20}$} (x30);
    \draw[e1] (x22) to node [right=-0.05,pos=0.5] {\hspace{.5mm}$\beta^{(3)}_{21}$} (x31);

    \draw[e0] (x31) to node [left=0.25,pos=0.3] {\hspace{.5mm}$\alpha^{(4)}_{0}$} (l);
    \draw[e0,bend right=30] (x30) to node [below left=-0.05,pos=0.25] {\hspace{.5mm}$\alpha^{(4)}_{1}$} (l);
    \draw[e1,bend left=30] (x31) to node [below right=-0.1,pos=0.3] {\hspace{.5mm}$\beta^{(4)}_{1}$} (l);
    \draw[e1] (x30) to node [right=0.25,pos=0.3] {\hspace{.5mm}$\beta^{(4)}_{0}$} (l);
\end{tikzpicture}};
    \node[anchor=south west] at ([xshift=+3.4cm,yshift=0.6cm]left_fig.north west) {\textbf{}};

    \node (right_fig) at (10,1.2) {\begin{tikzpicture}[auto, thick,node distance=3.cm,inner sep=1.5pt,scale=0.8]
    \newcommand\dist{.33cm}
    \newcommand\sep{.7cm}
    \newcommand\sepp{0.15cm}
    \newcommand\seppp{-.15cm}
    \node[] (0_l) [] {}; 

    \node[] (1_l) [below = 2*\dist of 0_l] {};
    \node[main] (1x10)  [left = 4*\dist of 1_l] {$0$};
    \node[main] (1x11)  [ right = 4*\dist of 1_l] {$1$};
    \node[] (1_r) [right = 2.8*\dist of 1x11] {};
    \node[] (1_m) [below = 1.5*\dist of 1_r] {\tikz \draw[dashed, <->] (-0.2,0) -- (0.4,0);};
    \node[] (1_m_r) [right= 0.1*\dist of 1_m] {$\mathcal{A}(2)[.,.,0]=\begin{pmatrix}
                    \alpha^{(2)}_{00} & \alpha^{(2)}_{01} &0 \\
                    0 &\alpha^{(2)}_{11} & \alpha^{(2)}_{12}
                \end{pmatrix}$};
    \node[main] (1x21) [below = 5*\dist of 1_l] {$1$};
    \node[main] (1x20) [left  = 6*\dist of 1x21] {$0$};
    \node[main] (1x22) [right  = 6*\dist of 1x21] {$2$};

    \node[] (2_l) [below = 3*\dist of 1x21] {};
    \node[main] (2x10)  [left = 4*\dist of 2_l] {$0$};
    \node[main] (2x11)  [ right = 4*\dist of 2_l] {$1$};
    \node[] (2_r) [right = 3.3*\dist of 2x11] {};
    \node[] (2_m) [below = 1.5*\dist of 2_r] {\tikz \draw[ <->] (-0.2,0) -- (0.4,0);};
    \node[] (2_m_r) [right= 0.1*\dist of 2_m] {$\mathcal{A}(2)[.,.,1]=\begin{pmatrix}
                    \beta^{(2)}_{00} & 0 &\beta^{(2)}_{02} \\
                    0 &0 & \beta^{(2)}_{12}
                \end{pmatrix}$};
    \node[main] (2x21) [below = 5*\dist of 2_l] {$1$};
    \node[main] (2x20) [left  = 6*\dist of 2x21] {$0$};
    \node[main] (2x22) [right  = 6*\dist of 2x21] {$2$};

    \draw[e0,bend right=30] (1x10) to node [left=0.1,pos=0.3] {\hspace{.5mm}$\alpha^{(2)}_{00}$} (1x20);
    \draw[e0,bend right=20] (1x10) to node [right=0.1,pos=0.5] {\hspace{.5mm}$\alpha^{(2)}_{01}$} (1x21);

    \draw[e0] (1x11) to node [left=0.05,pos=0.5] {\hspace{.5mm}$\alpha^{(2)}_{12}$} (1x22);
    \draw[e0] (1x11) to node [left=0.05,pos=0.2] {\hspace{.5mm}$\alpha^{(2)}_{11}$} (1x21);

    \draw[e1] (2x10) to node [right=-0.05,pos=0.5] {\hspace{.5mm}$\beta^{(2)}_{00}$} (2x20);
    \draw[e1] (2x10) to node [below=0.1,pos=0.7] {\hspace{.5mm}$\beta^{(2)}_{02}$} (2x22);

    \draw[e1,bend left=30] (2x11) to node [left=0.1,pos=0.5] {\hspace{.5mm}$\beta^{(2)}_{12}$} (2x22);

\end{tikzpicture}};
    \node[anchor=south west] at ([xshift=+5cm,yshift=0.6cm]right_fig.north west) {\textbf{}};

    \draw[<->, thick, dashed] (3.7, 1.7) -- (4.8, 2.1); %
    \draw[<->, thick] (3.7, 0.2) -- (4.8, -0.4); %

    \end{tikzpicture}
    \caption{A non-deterministic edge-valued decision diagram (nd-EVDD) and its equivalence to a tensor train (TT). A pseudo-Boolean function $f(x_1,\dots,x_4)$ is encoded as an nd-EVDD, which is structurally equivalent to a TT representation. The red edges highlight the bipartite subgraph between levels 2 and 3 of the diagram, whose adjacency matrix directly corresponds to one of the tensors in the TT. Dashed and black double arrows indicate how this bipartite graph decomposes into two subgraphs: one for low edges and one for high edges, encoding the matrices $\mathcal{A}_{0}^{(2)}$ and $\mathcal{A}_{1}^{(2)}$, respectively, each of dimension $2 \times 3$. In this construction, nodes at the upper level represent matrix rows, nodes at the lower level represent columns, and an edge between them corresponds to a non-zero matrix element, with the edge weight specifying its value; zero-valued elements are omitted for clarity. }
    \label{fig:nEVDD_example}
\end{figure*}

\subsection{From Tensor Train to nd-EVDD}
\label{sec:proof:MPS->EVDD}

Assume that we have a tensor train $\calA(1),\dots,\calA(n)$ and its corresponding dimensions $\chi_1,\dots,\chi_{n+1}$. The nd-EVDD that is equivalent to the given tensor train is defined as follows. For every $r\in[n+1]$, we define $\chi_r$ nodes $v^{(r)}_s$, with $s\in[\chi_r]$ labeled with variable $x_r$. The node $v^{(1)}_1$ will be the source of the EVDD, and $v^{(r+1)}_1$ is the sink of the EVDD, which is not labeled by a variable by definition. There are $b$-edges between nodes $v^{(r)}_s$ and $v^{(r+1)}_t$ with weight $\calA(r)[s,t,b]$, where $b\in\{0,1\},r\in[n],s\in[\chi_r],t\in[\chi_{r+1}]$.

We recursively define $\hat\calA(n)=\calA(n)$ and $\hat\calA(r)=\calA(r)\ast_{2,1}\hat\calA(r+1)$ for $r\in[n-1]$. Note that $\hat\calA(1)$ is the $2\times\dots\times2$ tensor that represents the function $f$.

We will show by induction on $r$ that the EVDD defined above represents the same function as the tensor train. We show that every node $v^{(r)}_s$ represents the function $f^{(s)}(\alpha)=\hat\calA(r)[s,\alpha]$ with $\alpha \in\{0,1\}^{n-r}$.

\paragraph{Base case} The node $v^{(n)}_s$ has $b$-edge with weight $\calA(n)[s,b]$ to the sink, and hence represents the function $f^{(s)}(x)=\calA(n)[s,x]$.

\paragraph{Induction case} Let $\alpha$ be an assignment and let $w_s=v^{(r)}_s$ be a node. We define $P_\alpha^{w_s}$ as the set of paths from node $w_s$ to the sink, following assignment $\alpha$. Then node $w_s$ represents the function
$f^{(s)}(\alpha)=\sum_{p\in P_\alpha^{w_s}}\prod_{e\in E(p)}w(e)$. Let $P_\alpha^{w,u}$ be the subset of paths from $P_\alpha^w$ that go via node $u$. Then we can rewrite with $u_t = v^{(r+1)}_t$
\begin{align*}
    f^{(s)}(\alpha)&=\sum_{t\in[\chi_{r+1}]}\sum_{p\in P_\alpha^{w_s,u_t}}\prod_{e\in E(p)}w(e)
    \\
    &=\sum_{t\in[\chi_{r+1}]}w(w_s\to^{\alpha(x_r)} u_t)\sum_{p\in P_\alpha^{u_t}}\prod_{e\in E(p)}w(e)\\
    &=\sum_{t\in[\chi_{r+1}]}w(w_s\to^{\alpha(x_r)} u_t)\cdot f^{(t)}(\alpha),
\end{align*}
where $w_s\to^b u_t$ is the $b$-edge from $w_s$ to $u_t$, and $f^{(t)}$ is the function represented by node $u_t$. By construction, we have $w(w_s\to^{\alpha(x_r)} u_t)=\calA(r)[s,t,\alpha(x_r)]$. By induction, we know that $f^{(t)}$ equals $\hat\calA(r+1)[t,\cdot]$. Hence, we have by definition of tensor contraction and of $\hat\calA(r)$
\begin{align*}
    f^{(s)}(\alpha) &= \sum_{t\in[\chi_{r+1}]} \calA(r)[s,t,\alpha(x_r)]\cdot \hat\calA(r+1)[t,\alpha] \\
    &= (\calA(r)\ast_{2,1}\hat\calA(r+1))[s,\alpha] =\hat\calA(r)[s,\alpha],
\end{align*}
which shows the induction.

Hence, the source $v^{1}_1$ represents the function $f=\hat\calA(1)$.

\begin{theorem}\label{theorem:TT->EVDD}
There is a procedure that, given a TT $\calA(1),\dots,\calA(n)$ representing the function $f\colon \{0,1\}^n \rightarrow \mathbb{K}$, constructs in linear time in the size of $\calA$ an nd-EVDD that represents the function $f$. 
\end{theorem}

\subsection{From nd-EVDD to Tensor Train}
\label{sec:proof:EVDD->MPS}

First, we assume that EVDDs are \emph{complete}, i.e., that on every path all variables from $X$ are read.  This can always be enforced in time linear in the size of the EVDD and $|X|$ as follows. On every edge connecting two nodes that skips $k$ intermediate variables, we place $k$ new nodes labeled with the intermediate variables, and pairs of 0-edges and 1-edges carrying weight $1$ that connect the newly introduced nodes to each other and to the two nodes of the original edge.

Now, we enumerate the nodes that belong to the same variable. Every node will be labeled as $v^{(r)}_s$, where $x_r$ is the variable of the node, and $s\in[\chi_r]$ is the enumerator. The sink will be labeled as $v^{(r+1)}_1$ We define the tensors $\calA(r)[s,t,b]$ as the weight of the $b$-edge connecting $v^{(r)}_s$ and $v^{(r+1)}_t$, for $r\in[n]$. This can be done in quadratic time, as there might be edges with weight 0.

We will show by induction that the tensor $\hat\calA(r)[s,\cdot]$ represents the same functions $f^{(s)}\colon\{0,1\}^{n-r}\to\mathbb{K}$ with $s\in[\chi_r]$ as nodes $\{v^{(r)}_s\}_{s\in[\chi_r]}$.

\paragraph{Base case.} For every assignment $\alpha$ there is only a single path from $v^{(n)}_s$ to sink. So $f^{(s)}$ takes values $f^{(s)}(\alpha)=\sum_{p \in P_\alpha}\prod_{e\in p}w(e)=w(v^{(r)}_s\to^{\alpha(x)}v^{(r+1)}_1)$. Tensor $\hat\calA(n)[s,\cdot]$ represents the functions $\hat\calA(n)[s,\alpha]=w(v^{(r)}_s\to^{\alpha(x_n)}v^{(r+1)}_1)=f^{(s)}(\alpha)$.

\paragraph{Induction case.}
Assume that tensor $\hat\calA(r+1)$ represents the functions $g^{(t)}$, corresponding to the functions represented by nodes $v^{(r+1)}_t$.
For every assignment $\alpha$, we see that the tensor $\hat\calA(r)$ can be decomposed as follows.
\begin{align*}
    \hat\calA(r)[s,\alpha]
    &=(\calA(r)\ast_{2,1}\hat\calA(r+1))[s,\alpha]\\
    &=\sum_{t\in[\chi_{r+1}]}\calA(r)[s,t,\alpha(x_r)]\cdot\hat\calA(r+1)[t,\alpha]\\
    &=\sum_{t\in[\chi_{r+1}]}w(v^{(r)}_s\to^{\alpha(x_{r})}v^{(r+1)}_t)\cdot g^{(t)}(\alpha)
\end{align*}
We can decompose $g^{(t)}(\alpha)$ by their definitions as functions representing nodes $v^{(r+1)}_t$. Thus we can rewrite $\hat\calA(r)[s,\alpha]$ as 
\begin{align*}
    &\sum_{t\in[\chi_{r+1}]}w(w_s\to^{\alpha(x_r)} u_t)\sum_{p\in P_\alpha^{u_t}}\prod_{e\in E(p)}w(e)\\
    &=\sum_{t\in[\chi_{r+1}]}\sum_{p\in P_\alpha^{w_s,u_t}}\prod_{e\in E(p)}w(e)\\
    &=\sum_{p\in P^{v^{(r)}_s}_\alpha}\prod_{e\in E(p)}w(e),
\end{align*}
i.e., exactly the functions represented by nodes $\{v^{(r)}_s\}_{s\in[\chi_r]}$.
Hence, $\hat\calA(1)$ represents the same function as the source node $v^{(1)}_1$, so the constructed tensor train represents the same function as the given EVDD.

\begin{theorem}\label{theorem:EVDD->TT}
There is a procedure that, given a nd-EVDD $G$ representing the function $f\colon \{0,1\}^n \rightarrow \mathbb{K}$, constructs in quadratic time in the size of $G$ and linear time in $n$ a tensor train $\calA(1),\dots,\calA(n)$ that represents the function $f$. 
\end{theorem}

\subsection{Deterministic EVDD}

From the construction of the tensor train in Section~\ref{sec:proof:EVDD->MPS}, we directly observe the relation between deterministic EVDDs and tensor trains. For deterministic EVDDs, there is at most one outgoing 0-edge and at most one outgoing 1-edge. Therefore, in the corresponding tensor train, the tensors $\calA(r)[s,t,i]$ have at most one non-zero value in the vector $\calA(r)[s,\cdot,i]$ for $s\in\chi_r,i\in\{0,1\}$. In other words, all matrices $\calA(r)[\cdot,\cdot,i]$ have at most one non-zero entry per row. By the construction of an EVDD from a TT in Section~\ref{sec:proof:MPS->EVDD}, we see that tensor trains with this property always correspond to a deterministic EVDD. 

\cite{vinkhuijzen2024knowledge} showed that a TT is always at least as succinct as a deterministic EVDD up to a polynomial factor, but there exist families of states for which TT is exponentially more succinct than deterministic EVDD. This corresponds to our observation that deterministic EVDDs correspond to a strict subset of tensor trains.

\section{From Tree Tensor Networks to Structured-Decomposability, and back}

We explain the equivalence of tree tensor networks and structured-decomposable $(+,\times)$-circuits by showing two transformations. Figure~\ref{fig:TTN-SDD} shows an example of a tree tensor network and an equivalent structured-decomposable circuit.

\subsection{From TTN to Circuits}
\label{subsec:TTN->circuit}Let $(\calA, T, d)$ be a TTN. The vtree of the corresponding structured-decomposable circuit is $\calT = (T,\lambda)$, where $T$ is the underlying tree of the TTN and $\lambda$ maps its $n$ leaves, read from left to right, to the variables $x_1,\dots, x_n$. For $v \in V(T)$, we denote by $T_v$ the tree below $v$ in $T$.
We explain the translation inductively.

Let $v$ be a node of $T$. Recall that if $\hat \calA(v)$ is a $2 \times \dots \times 2 \times d(v)$ tensor over $\mathbb{K}$, with $m$ $2$s corresponding to the $m$ leaves below $v$, then we interpret each $\hat\calA(v)[\cdot,k]$ for $k\in[d(v)]$ as a pseudo-Boolean function in $\{0,1\}^m \rightarrow \bbK$. So $\hat \calA(v)$ corresponds to a sequence of $d(v)$ pseudo-Boolean functions.

\paragraph*{Terminal node.} If $v$ is a leaf of $T$ then $\hat\calA(v) = \calA(v)$ is a $2 \times d(v)$ tensor over $\mathbb{K}$. Let $x$ be the variable $\lambda(v)$ corresponding to $v$. We have 
\[
\hat\calA(v)[x,i] = \calA(v)[0,i] \cdot \mathbf{1}[x = 0] + \calA(v)[1,i] \cdot \mathbf{1}[x = 1]
\]
where, $\mathbf{1}[x = b]$ is the indicator function that returns $1$ when $x$ is set to $b$ and $0$ otherwise. We then define the gates $g^{1}_v,\dots,g^{d(v)}_v$ as follows
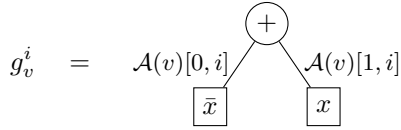
\[
g^{i}_v \quad = \quad \begin{tikzpicture}[xscale=1.5,yscale=1.1,baseline=(l.base)]
\node[draw,minimum height=0.5cm] (nx) at (0,0) {$\bar x$};
\node[draw,minimum height=0.5cm] (x) at (1,0) {$x$};
\node[circle,draw,inner sep=2pt] (n) at (0.5,1) {$+$};
\draw (x) to node[midway,right,font=\footnotesize] (l) {$\calA(v)[1,i]$} (n);
\draw (nx) to node[midway,left,font=\footnotesize] {$\calA(v)[0,i]$} (n);
\end{tikzpicture}
\]
It is readily verified that $g^i_v$ computes $\hat\calA(v)[\cdot,i]$ and that it is structured by the vtree made of a single leaf labeled with~$x$. 

\paragraph*{Internal node.} Let $v$ be an internal node of $T$ with left and right children $v_\ell$ and $v_r$. Then we have 
$
\hat\calA(v) = \calA(v) \ast \hat\calA(v_\ell)  \ast \hat\calA(v_r)
$.
To improve readability we write $d_\ell = d(v_\ell)$ and $d_r = d(v_r)$. By inductive assumption, we have already constructed the circuits below the gates $g^{i}_{v_\ell}$ and $g^{j}_{v_r}$ for every $i \in [d_\ell]$ and $j \in [d_r]$. Let $n = n_\ell + n_r$ be the number of leaves below $v$, with $n_\ell$ below $v_\ell$ and $n_r$ below $n_r$. For $k \in [d(v)]$, the $k^\text{th}$ function $\hat \calA(v)[\cdot,k]$ is 
\vspace{-1em}
\begin{align*}
\hat\calA(v)[x_1,...,x_n,k] = \quad
\mathclap{\sum_{\substack{k_\ell,k_r \\ \in [d_\ell]\times [d_r]}}}\quad \begin{array}{r}
 \text{ } \\   \text{ } \\ \hat\calA(v_r)[x_{n_\ell+1},...,x_n,k_r] \\
 \cdot \hat\calA(v_\ell)[x_1,...,x_{n_\ell},k_\ell]
 \\ \cdot \calA(v)[k_\ell,k_r,k] 
\end{array}\tag{1}
\end{align*}
We define the gates $g^{1}_v,\dots,g^{d(v)}_v$ as follows. $g^k_v = $
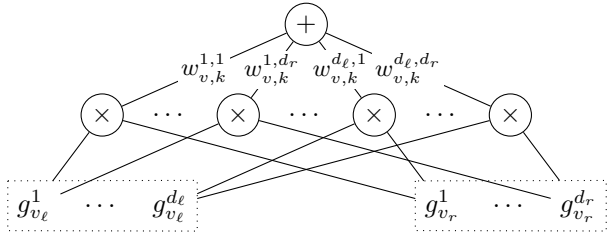
\[
\begin{tikzpicture}[xscale=1.8,yscale=1.2,baseline=(l.base)]
\node[circle,draw,inner sep=2pt] (x1) at (0,0) {$\times$};
\node (dots) at (-0.5,0) {$\dots$};
\node[circle,draw,inner sep=2pt] (x2) at (1,0) {$\times$};
\node (dots) at (1.5,0) {$\dots$};
\node (dots) at (0.5,0) {$\dots$};
\node[circle,draw,inner sep=2pt] (x3) at (2,0) {$\times$};
\node[circle,draw,inner sep=2pt] (x4) at (-1,0) {$\times$};
\node[circle,draw,inner sep=2pt] (n) at (0.5,1) {$+$};
\draw (x1) to node[midway,fill=white,inner sep=1pt,font=\footnotesize] (l) {$w^{1,d_r}_{v,k}$} (n);
\draw (x2) to node[midway,fill=white,inner sep=1pt,font=\footnotesize] {$w^{d_\ell,1}_{v,k}$} (n);
\draw (x3) to node[midway,fill=white,inner sep=1pt,font=\footnotesize] (l) {$w^{d_\ell,d_r}_{v,k}$} (n);
\draw (x4) to node[midway,fill=white,inner sep=1pt,font=\footnotesize] {$w^{1,1}_{v,k}$} (n);
\node (a) at (2.5,-1) {$g^{d_r}_{v_r}$};
\node (dots) at (2,-1) {$\dots$};
\node (b) at (1.5,-1) {$g^{1}_{v_r}$};
\node (c) at (-0.5,-1) {$g^{d_\ell}_{v_\ell}$};
\node (dots) at (-1,-1) {$\dots$};
\node (d) at (-1.5,-1) {$g^{1}_{v_\ell}$};
\draw (c) -- (x3) -- (a);
\draw (c) -- (x2) -- (b);
\draw (d) -- (x1) -- (a);
\draw (d) -- (x4) -- (b);

\draw[dotted] (b.north west) -- (b.south west) -- (a.south east) -- (a.north east) -- cycle;

\draw[dotted] (d.north west) -- (d.south west) -- (c.south east) -- (c.north east) -- cycle;
\end{tikzpicture}
\]
where $w^{i,j}_{v,k} = \calA(v)[i,j,k]$.
 It follows from the above expression of $\hat\calA(v)[\cdot,k]$ that, if for every $i,j \in [d_\ell] \times [d_r]$, $g^i_{v_\ell}$ computes $\hat\calA(v_\ell)[\cdot,i]$ and $g^j_{v_r}$ computes $\hat\calA(v_r)[\cdot,j]$, then $g^k_v$ computes $\hat\calA(v)[\cdot,k]$. By induction, every circuit in the left box is structured by the vtree $\calT_{v_\ell}$ and every circuit in the right box is structured by the vtree $\calT_{v_r}$. Thus, the circuit under $g^k_v$ is structured by $T_v$. One counts that the construction of $g^k_v$ introduces $1 + d_\ell d_r$ new gates. The same $d_\ell d_r$ $\times$-gates are reused by all $g^1_v,\dots,g^{d(v)}_v$ and thus constructing all these gates requires $d(v) + d_rd_l = d(v) + d(v_\ell)d(v_r)$ new gates and $(2+d(v))d(v_\ell)d(v_r)$ new edges in total. Therefore, assuming the circuits in the boxes are accessed in $O(1)$ time, constructing  $g^1_v,\dots,g^{d(v)}_v$ takes time linear in the size of $\calA(v)$.

Thus, this TTN-to-circuit translation is a linear-time construction. If $(\calA,T,d)$ represents a single function, i.e., if $\hat \calA$ is a $2^n \times 1$ tensor, then the final circuit has a single output gate representing this exact function. Otherwise, if the TTN represents a sequence of $s$ functions, then the final circuit has exactly $s$ output gates, one per function.

\begin{theorem}\label{theorem:TTN->circuit}
There is a procedure that, given a TTN $(\calA,T,d)$ representing the functions $f^1,\dots,f^s \colon \{0,1\}^n \rightarrow \mathbb{K}$, constructs in linear time in the size of $\calA$ an $s$-output structured-decomposable circuit such that, for every $k \in [s]$, the $k^\text{th}$  output gate represents~$f^k$. 

For every vtree node $v$, the circuit has exactly $d(v)$ $+$ gates $g^1_v,\dots,g^{d(v)}_v$ mapped to $v$ and $g^k_v$ computes exactly the function $\hat\calA(v)[\cdot,k]$.
\end{theorem}

\subsection{From Circuits to TTN}\label{section:circuits->TTN}

The backward translation from structured-decomposable circuits to TTN is basically the reverse of the forward translation explained above but the circuit has to go through some (simple) preprocessing. Let $C$ be an $(S,+,\times,0,1)$ circuit structured by $\calT = (T,\lambda)$ over the variables $x_1,\dots,x_n$ and let $\phi$ be the mapping between $C$'s gate and $T$'s node. We modify $C$ (while preserving the function computed) so that it respects the following properties: 
\begin{itemize}
\item[1.] The output gates of $C$ are all $+$ gates.
\item[2.] For each $+$ gate $g$, the children of $g$ are either circuit inputs or $\times$ gate.
\item[3.] There is at most one input gate $\mathbf{1}[x_i = b]$ per $i \in [n]$ and $b \in \{0,1\}$.
\end{itemize}
The first two properties are easily enforced in linear time by interleaving a dummy $+$ gate with a single incoming edge carrying the unit weight at every edge of $C$ that connects a $\times$ gate to another $\times$ gate or to a circuit input. The third property is also implemented in linear time by merging identical inputs. Next, for every $v \in V(T)$, we fix an arbitrary total order upon the $+$ gates mapped to $v$ by $\phi$. We denote the resulting gate sequence by $g^1_v,\dots,g^{d_v}_v$. We modify $C$ so that it becomes \emph{fully-connected} in the following sense:
\begin{itemize}
\item[4.] For every $g^k_v$, if $v$ is the leaf mapped to $x$ by $\lambda$, then $g^k_v$ is of the form $w^0_{v,k} \cdot \mathbf{1}(x = 0) +  w^1_{v,k} \cdot \mathbf{1}(x = 1)$ for some weights $w^0_{v,k},w^1_{v,k} \in S$; if however $v$ is an internal node with left and right children $v_\ell$ and $v_r$, then $g^k_v = \sum_{i,j \in [d_{v_\ell}] \times [d_{v_r}]} w^{i,j}_{v,k} \cdot (g^i_{v_\ell} \cdot g^j_{v_r})$ with $w^{i,j}_{v,k} \in S$. 
\end{itemize}
The fully-connected property is implemented in quadratic time by adding every missing incoming edge to $+$ gates with $0$ weight.

Now we can define the tensor $\calA(v)$. If $v$ is a leaf node then $\calA(v)$  is the $2 \times d_v$ tensor defined by $\calA(v)[0,k] = w^0_{v,k}$ and $\calA(v)[1,k] = w^1_{v,k}$. If $v$ is an internal node then $\calA(v)$ is the $d_{v_\ell} \times d_{v_r} \times d_v$ tensor defined by $\calA(v)[i,j,k] = w^{i,j}_{v,k}$. The final tree tensor network is $(\calA,T,d)$, with $d$ mapping each $v \in V(T)$ to $d_v$. If $C$ respects the four properties 1., 2., 3., and 4., then the TTN is built in linear time.

Transforming the TTN back to a structured-decomposable circuit as in Section~\ref{subsec:TTN->circuit} gives back $C$, hence the following reverse statement to Theorem~\ref{theorem:TTN->circuit}.

\begin{theorem}\label{theorem:circuit->TTN}
There is a procedure that, given an $s$-outputs structured-decomposable circuit $C$ structured by $((T,\lambda),\phi)$ and representing $f^1,\dots,f^s \colon \{0,1\}^n \rightarrow \mathbb{K}$, processes $C$ in quadratic time into an equivalent circuit $C'$ also structured by $(T,\lambda)$ and then turns $C'$ in linear time into a TTN $(\calA,T,d)$ representing $f^1,\dots,f^s$. 

In addition, for every $v \in V(T)$, if $C'$ has $d_v$ $+$ gates $g^1_v,\dots,g^{d_{v}}_v$ mapped to $v$ by $\phi$, then $d(v) = d_v$ and $\hat\calA(v)[\cdot,k]$ exactly compute $g^k_v$ for every $k \in [d_v]$. 
\end{theorem}

The whole procedure described in Theorem~\ref{theorem:circuit->TTN} takes quadratic time, while that of turning TTN into circuits takes linear time. This difference is purely due to the fact that, when constructing the TTN, we populate the tensors with many $0$s corresponding to missing edges in the circuits. With a more compact representation of the tensors in the TTN, for instance, storing only a list of non-zero entries, the transformation can be done more quickly.

\section{Determinism}

Determinism is a crucial property of circuits in many applications. Here we investigate what this property becomes in the world of TTN.

\subsection{Decision Determinism }

To enforce determinism in practice, one option is to have all $+$ gates of the form 
$$
g_0 \cdot \mathbf{1}[x = 0] + g_1 \cdot \mathbf{1}[x = 1]
$$
for some variable $x$ and circuits $g_0$ and $g_1$. Circuits that are structured-decomposable and respect the above property are called \emph{decision structured-decomposable}: at gate $g$, we \emph{decide} whether to evaluate $g_0$ or $g_1$ based on the value of $x$. We also say that the gate \emph{branches} on $x$.

From a theoretical viewpoint, one disadvantage of these circuits is that, while every function can be represented as a decision structured-decomposable circuit for \emph{some} vtree, not all functions can be represented as decision structured-decomposable circuits for \emph{all} vtrees. For example, consider the function $x_1 + x_2 + x_3 + x_4$ in the semi ring of integers and the vtree
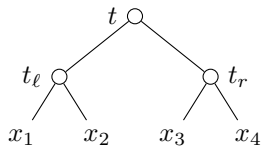
\begin{center}
\begin{tikzpicture}[yscale=0.8]
\node[draw, circle,inner sep = 2,label={west:$t$}] (t) at (0,0) {};
\node[draw, circle,inner sep = 2,label={west:$t_\ell$}] (l) at (-1,-1) {};
\node[draw, circle,inner sep = 2,label={east:$t_r$}] (r) at (+1,-1) {};
\node (a) at (-1.5,-2) {$x_1$};
\node (b) at (-0.5,-2) {$x_2$};
\node (c) at (0.5,-2) {$x_3$};
\node (d) at (1.5,-2) {$x_4$};

\draw (l) -- (t) -- (r);
\draw (a) -- (l) -- (b);
\draw (c) -- (r) -- (d);
\end{tikzpicture}
\end{center}
In a decision structured-decomposable circuit respecting this vtree, $+$ gates can only be mapped to $t_\ell$ and $t_r$, whereas gates mapped to $t$ must be decomposable products. Thus, to get such circuit for $x_1 + x_2 + x_3 + x_4$ we would have to write $x_1 + x_2 + x_3 + x_4$ as a product of functions $f_\ell(x_1,x_2)\cdot f_r(x_3,x_4)$. But this is not possible since we would need $f_\ell(0,0)\cdot f_r(0,0) = 0$, which would imply that $f_\ell(0,0) = 0$ or $f_r(0,0) = 0$, which in turn would imply 
that $f_\ell(1,1)\cdot f_r(0,0) = 0$ or $f_\ell(0,0)\cdot f_r(1,1) = 0$, while the correct answer is $2$.

Decision structured-decomposable circuits are equivalent to TTNs that respect a local property, i.e., a property that is respected at every node $v$ of the tree by the tensor $\calA(v)$. The functions $f^1_v, \dots, f^{d(v)}_v$ can only be of one type, which only depends on the position of $v$ in the tree. 

\paragraph*{Terminal node.} If $v$ is  a leaf of $T$ for the variable $x$ then each $f^i_v$ is a function computed by a circuit of the form
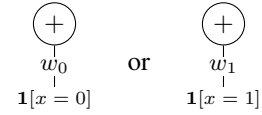
\[
\begin{tikzpicture}[xscale=1,yscale=1,baseline=(l.base)]

\node[font=\scriptsize,inner sep=1pt] (nx) at (-0.5,-2) {$\mathbf{1}[x = 0]$};

\node[circle,draw,inner sep=2pt] (c) at (-0.5,-1) {$+$};

\draw (c) to node[midway,fill=white,inner sep=1pt,font=\footnotesize] (l) {$w_0$} (nx);
\end{tikzpicture}
\quad \text{ or } \quad 
\begin{tikzpicture}[xscale=1,yscale=1,baseline=(l.base)]

\node[font=\scriptsize,inner sep=1pt] (nx) at (-0.5,-2) {$\mathbf{1}[x = 1]$};

\node[circle,draw,inner sep=2pt] (c) at (-0.5,-1) {$+$};

\draw (c) to node[midway,fill=white,inner sep=1pt,font=\footnotesize] (l) {$w_1$} (nx);
\end{tikzpicture}
\]
So $\calA(v)$ is a $2 \times d(v)$ tensor such that, for every $i \in [d(v)]$,
$$
\calA(v)[0,i] = 0\quad \text{or}\quad \calA(v)[1,i] = 0
$$

\paragraph*{Almost terminal node.} If $v$ is an internal node with one of its children that is a leaf, say $v_r$, then each $f^i_v$ is a function computed by a circuit of the form

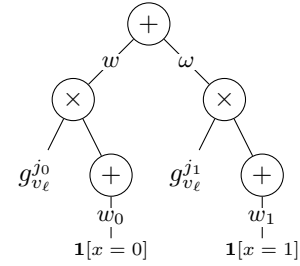
\[
\begin{tikzpicture}[xscale=1,yscale=1,baseline=(l.base)]
\node[circle,draw,inner sep=2pt] (x1) at (-1,0) {$\times$};
\node[circle,draw,inner sep=2pt] (x2) at (+1,0) {$\times$};
\node[circle,draw,inner sep=2pt] (n) at (0,1) {$+$};
\draw (x1) to node[midway,fill=white,inner sep=1pt,font=\footnotesize] (l) {$w$} (n);
\draw (x2) to node[midway,fill=white,inner sep=1pt,font=\footnotesize] (l) {$\omega$} (n);
\node (a) at (-1.5,-1) {$g^{j_0}_{v_\ell}$};
\node (d) at (0.5,-1) {$g^{j_1}_{v_\ell}$};

\node[font=\scriptsize,inner sep=1pt] (nx) at (-0.5,-2) {$\mathbf{1}[x = 0]$};
\node[font=\scriptsize,inner sep=1pt] (x) at (1.5,-2) {$\mathbf{1}[x = 1]$};

\node[circle,draw,inner sep=2pt] (b) at (1.5,-1) {$+$};
\node[circle,draw,inner sep=2pt] (c) at (-0.5,-1) {$+$};
\draw (c) -- (x1) -- (a);
\draw (b) -- (x2) -- (d);
\draw (c) to node[midway,fill=white,inner sep=1pt,font=\footnotesize] (l) {$w_0$} (nx);
\draw (b) to node[midway,fill=white,inner sep=1pt,font=\footnotesize] (l) {$w_1$} (x);
\end{tikzpicture}
\]
where $x$ is the variable for $v_r$. So the matrix $\calA(v)[\cdot,\cdot,i]$ has only two entries that are not $0$, that respectively correspond to the $\times$ gates: there are at most two indexes $j_0,j_1 \in [d(v_\ell)]$ and at most two indexes $k_0,k_1 \in [d(v_r)]$ such that 
$$
\calA(v)[j_0,k_0,i] \neq 0 \quad \text{and}\quad \calA(v)[j_1,k_1,i] \neq 0,
$$
furthermore, $f^{k_0}_{v_r} = w_0 \cdot \mathbf{1}[x = 0]$ and  $f^{k_1}_{v_r} = w_1 \cdot \mathbf{1}[x = 1]$. The case where $v_\ell$ is a leaf of $T$ is analogous.

\paragraph*{Internal node.} If $v$ is an internal node of $T$ whose children are not leaves of $T$ then each $f^i_v$ is a function computed by a circuit of the form
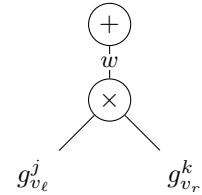
\[
\begin{tikzpicture}[xscale=1,yscale=1,baseline=(l.base)]
\def\o{4};
\node[circle,draw,inner sep=2pt] (x1) at (0,0) {$\times$};
\node[circle,draw,inner sep=2pt] (n) at (0,1) {$+$};
\draw (x1) to node[midway,fill=white,inner sep=1pt,font=\footnotesize] (l) {$w$} (n);
\node (a) at (+1,-1) {$g^{k}_{v_r}$};
\node (d) at (-1,-1) {$g^{j}_{v_\ell}$};
\draw (d) -- (x1) -- (a);
\end{tikzpicture}
\]
So the matrix $\calA(v)[\cdot,\cdot,i]$ has only one entry that is not $0$, which corresponds to the $\times$ gate: there is only one pair of index $j,k \in [d(v_\ell)] \times [d(v_r)]$ such that 
$
\calA(v)[j,k,i] \neq 0.
$

Thus, decision structured-decomposable circuits correspond to TTNs with very sparse tensors. We call \emph{decision TTNs} the TTN whose tensors $\calA(v)$ are of the type described above. We then have the following equivalence.

\begin{theorem}
There is a procedure that, given an $s$-outputs decision structured-decomposable circuit representing $f^1,\dots,f^s \colon \{0,1\}^n \rightarrow \mathbb{K}$, constructs in polynomial time a decision TTN representing the same $s$ functions. 
\end{theorem}

\begin{theorem}
There is a procedure that, given a decision TTN representing $f^1,\dots,f^s \colon \{0,1\}^n \rightarrow \mathbb{K}$, constructs in polynomial time an $s$-outputs decision structured-decomposable circuit representing the same $s$ functions.
\end{theorem}

\subsection{General Determinism}

Restricting to decision gates is not the only way to enforce determinism. In this section, we discuss what general determinism corresponds to in TTN.

On structured-decomposable circuits, checking that a $+$ gate $w_1\cdot g_1 + w_2 \cdot g_2$ is deterministic is doable in quadratic time: compute a structured-decomposable circuit representing $g_1 \times g_2$ and check whether its support is empty. The product operation runs in quadratic time because $g_1$ and $g_2$ are structured similarly. Checking that a gate is a decision gate is done locally by looking at its children, but checking general determinism seems to involve operations on all descendants of the $+$ gate. The same is to be expected for the deterministic version of TTN.

\begin{definition} A TTN $(\calA,T,d)$ is \emph{deterministic} when, for every $v \in V(T)$, every $k \in [d(v)]$ and every two distinct entries of $\calA(v)[i_1,j_1,k]$ and $\calA(v)[i_2,j_2,k]$ of $\calA(v)[\cdot,\cdot,k]$, if both entries are non-zero then $\hat\calA(v_\ell)[\cdot,i_1] \odot \hat\calA(v_\ell)[\cdot,i_2] = 0$ or $\hat\calA(v_r)[\cdot,j_1] \odot \hat\calA(v_r)[\cdot,j_2] = 0$.
\end{definition}

Recall that $\odot$ is the element-wise product of tensors. We claim that the procedures described in Theorems~\ref{theorem:TTN->circuit} and~\ref{theorem:circuit->TTN} turn deterministic circuits into deterministic TTN and vice versa. 

\paragraph*{From Circuits to TTN.}
In Theorem~\ref{theorem:circuit->TTN}, the processing from $C$ to $C'$ adds single-input $+$ gates and adds incoming edges to $+$ gates carrying weight $0$. So if $C$ is deterministic, then $C'$ is too. The procedure then constructs for every $+$ gate $g^k_v = \sum_{i,j \in [d_{v_\ell}] \times [d_{v_r}]} w^{i,j}_{v,k} \cdot g^i_{v_\ell} \cdot g^j_{v_r}$ of $C'$ a matrix $\calA(v)[\cdot,\cdot,k]$ such that $\calA(v)[i,j,k] = w^{i,j}_{v,k}$. 

When $C'$ is deterministic, for each $(i_1,j_1) \neq (i_2,j_2)$ with $w^{i_1,j_1}_{v,k} \neq 0$ and $w^{i_2,j_2}_{v,k} \neq 0$ we have $$g^{i_1}_{v_\ell} \cdot g^{j_1}_{v_r} \cdot g^{i_2}_{v_\ell} \cdot g^{j_2}_{v_r} = 0$$ 
Since $\var(g^{i_1}_{v_\ell} \cdot g^{i_2}_{v_\ell}) \cap \var(g^{j_1}_{v_r} \cdot g^{j_2}_{v_r}) \subseteq \var(v_\ell) \cap \var(v_r) = \emptyset$ we deduce that 
$$
g^{i_1}_{v_\ell} \cdot g^{i_2}_{v_\ell} = 0\quad \text{ or } \quad g^{j_1}_{v_r} \cdot g^{j_2}_{v_r} = 0
$$
By Theorem~\ref{theorem:circuit->TTN}, for every assignment $\alpha$ to $\var(v_\ell)$ we have $\hat\calA(v_\ell)[\alpha,i_1] = g^{i_1}_{v_\ell}(\alpha)$, $\hat\calA(v_\ell)[\alpha,i_2] = g^{i_2}_{v_\ell}(\alpha)$ and for every assignment $\beta$ to $\var(v_r)$ we have $\hat\calA(v_r)[\beta,j_1] = g^{j_1}_{v_r}(\beta)$ and $\hat\calA(v_r)[\beta,j_2] = g^{j_2}_{v_r}(\beta)$. Thus we indeed have that 
\begin{align*}
&\hat\calA(v_\ell)[\cdot,i_1] \odot  \hat\calA(v_\ell)[\cdot,i_2] = 0
\\
 \text{ or } 
 &\hat\calA(v_r)[\cdot,j_1] \odot  \hat\calA(v_r)[\cdot,j_2] = 0
\end{align*}

\paragraph*{From TTN to circuit.} 
The procedure of Theorem~\ref{theorem:TTN->circuit} turns the TTN $(\calA,T,d)$ into a structured-decomposable circuit respecting vtree $((T,\lambda),\phi)$ where each gate $g^k_v$ mapped to $v$ by $\phi$ computes exactly $\hat\calA(v)[\cdot,k]$. We then have  $
g^k_v = \hat\calA(v) $
\begin{align*}
 &= \sum_{k_\ell,k_r \in [d_\ell]\times [d_r]}\calA(v)[k_\ell,k_r,k] \cdot \hat\calA(v_\ell)[\cdot,k_\ell] \cdot \hat\calA(v_r)[\cdot,k_r] 
 \\
 &= \sum_{k_\ell,k_r \in [d_\ell]\times [d_r]}\calA(v)[k_\ell,k_r,k] \cdot g^{k_\ell}_{v_\ell}\cdot g^{k_r}_{v_r}
\end{align*}
If the TTN is deterministic, then $\calA(v)[k_\ell,k_r,k] \neq 0$ and $\calA(v)[h_\ell,h_r,k] \neq 0$ and $(k_\ell,k_r) \neq (h_\ell,h_r)$ forces $\hat\calA(v_\ell)[\cdot,k_\ell] 
\cdot \hat\calA(v_\ell)[\cdot,h_\ell] = 0$ or $\hat\calA(v_r)[\cdot,k_r] 
\cdot \hat\calA(v_r)[\cdot,h_r] = 0$ and therefore $g^{k_\ell}_{v_\ell} \cdot g^{h_\ell}_{v_\ell} = 0$ or $g^{k_r}_{v_r} \cdot g^{h_r}_{v_r} = 0$. It follows that $g^k_v$ is indeed a deterministic $+$ gates.

\begin{lemma}
The procedure of Theorem~\ref{theorem:TTN->circuit} turns deterministic TTN into deterministic structured-decomposable circuits.
\end{lemma}

\begin{lemma}
The procedure of Theorem~\ref{theorem:circuit->TTN} turns deterministic structured-decomposable circuits into deterministic TTN.
\end{lemma}

\section{Application to Representations for Quantum States}

Tensor networks and structured circuits are popular ways to describe quantum states. In this section, we explain how these representations are used in quantum computing.

\subsection{Quantum States}
Quantum states are described by a complex vector, called the quantum state vector. A quantum state consisting of $n$ qubits is represented by a vector of size $2^n$. In quantum computing and quantum mechanics, quantum state vectors are usually written in Dirac notation. In Dirac notation, a vector $\vec{v}$ is written as $\ket{\phi}$ (or with another Greek letter around the brackets $\ket{.}$). The basis vector $\vec{e_x}=(0,0,\dots,0,1,0,\dots,0)^T$, with a 1 at the $x$'th position and all 0's elsewhere, is written as $\ket{x}$, with $x\in\{0,1,\dots,2^n-1\}$, or its equivalent Boolean representation $x\in\{0,1\}^n$.

This notation directly relates pseudo-Boolean functions to vectors: every quantum state vector can be written as
$$\ket{\phi}=\sum_{\vec{x}\in\{0,1\}^n}f(\vec{x})\ket{\vec{x}},$$
where $f$ is a pseudo-Boolean function of the form $f\colon\{0,1\}^n\to\bbC$. Therefore, every quantum state, which is represented by a quantum state vector, can also be represented by a pseudo-Boolean function. Because every pseudo-Boolean function can be represented by a tensor network or structured circuit, a quantum state can, in fact, also be represented by those representations.

\subsection{Tensor Trains in Quantum Computing}
Recall that a TT is defined as an ordered array of $n$ tensors of order three, denoted by $\calA(r)$, where $r$ indicates the position of each tensor within the array. The function interpretation of a TT can be directly translated to a quantum state representation, yielding the state
$$
    \ket{\phi}=\sum_{x_{1},\dots, x_n\in\{0,1\}^n}\mathcal{A}{(1)}[x_1] \cdots\mathcal{A}{(n)}[x_n]\ket{x_{1},\dots, x_n},
$$
where we omitted the $\ast_{2,1}$ contractions between the tensors, as they are natural multiplications of the corresponding matrices.

\subsection{EVDD in Quantum Computing}
Recall that an EVDD is a graph with weighted edges. Its function interpretation for an assignment is the sum of products over paths from root to sink corresponding to the assignment. This may appear abstract or difficult to interpret in the context of quantum states, but there seems to be a fairly clean reformulation in Dirac notation. 

First, we observe that every node in an EVDD represents a function $f\colon\{0,1\}^k\to\bbK$ itself. Therefore, as every function is related to a vector, every node in the EVDD represents a vector over $\bbK$ of length $2^k$. We will write the vector represented by node $v$ as $\ket{\phi_v}$. Now, we can recursively write for every internal node $v$ of the EVDD
\begin{align*}
    \ket{\phi_v}
    =&~~~~\ket{0}\otimes\sum_{u \text{ a 0-child of }v}w(v\to^0 u)\ket{\phi_u} \\
    &+ \ket{1}\otimes\sum_{u \text{ a 1-child of }v}w(v\to^1 u)\ket{\phi_u},
\end{align*}
and the sink node is defined as $\ket{\phi_{sink}}=1$. Here, we used the Kronecker product $\otimes$, which acts on basis vectors as $\ket{x}\otimes\ket{y}=\ket{x,y}$ for any $x\in\{0,1\}^k$ and $y\in\{0,1\}^\ell$.

\section{Related Work}

Tensor networks originate in quantum many-body physics, and have since been applied to combinatorial optimization, numerical partial differential equations, and machine learning~\cite{Cirac_PEPS,Chamon_Search_Algorithm_MPS,Garcia_3SAT_TN,Boris_PDE_TN,Li_WAutomata_TN}. These representations compactly encode high-dimensional tensors while supporting efficient contraction algorithms.
Beyond tree-like topologies, cyclic variants like Projected Entangled Pair States (PEPS)~\cite{Cirac_PEPS} have been considered.

On the circuit side, other classes of circuits have been studied. For instance, decomposable circuits that do not implement \emph{structured-decomposability} are generally much more compact than the circuits considered in the present work~\cite{pipatsrisawat2008new}, but we fail to see their counterpart in the tensor network world.
Among structured-decomposable circuits, sentential decision diagrams (SDDs)~\cite{darwiche2011sdd,Kisa_PSDD} refine structured decomposability via canonicity relative to a vtree. SDDs have a TTN equivalent, and canonical SDDs should yield some kind of canonical TTN. But canonical SDDs themselves are rather unsatisfactory since, in order to attain canonicity, one has to give up the tractability of several operations (like conditioning or computing conjunction)~\cite{BroeckD15}. Similar drawbacks are expected for their TTN equivalent.

Tensor-based methods have been investigated for model counting~\cite{kourtis2019fast,DudekDV19,DudekPV20}, an area closely related to proposition decomposable and deterministic circuits (in particular, decision-DNNF and d-DNNF circuits). Tensor-based proof systems for model counting have recently been investigated ~\cite{BeyersdorffGGHKS26} and compared to existing proof systems, which, for most of them, construct tractable circuits. Compared to our work, tensor-based methods for model counting do not aim at computing a representation of the formula, but only its model count. 

Among recent research relating circuits and tensor networks we want to mention ~\cite{LoconteMGPCQVV25,DBLP:journals/corr/abs-2512-17090} where the connection between TTN and tractable probabilistic circuits is investigated in one direction (namely, from TTN to circuits).  We should also mention~\cite{Onaka0NY25a}, which uses tensor trains (TT) to represent Boolean functions but only considers deterministic TT. One small difference is that their TT use entries $1$, $0$ and $-1$ while ours would only use $1$ and $0$ in the Boolean semi-ring. 

Connections between decision diagrams and quantum state representations have been explored in quantum circuit simulation~\cite{Miller_DD_Qsim}. Our equivalences between tensor trains and non-deterministic EVDDs, and between TTNs and structured-decomposable circuits, formalize and generalize this relationship. More broadly, our work contributes to recent efforts to understand tractable representations from an algebraic perspective, including settings with negative or complex weights~\cite{valiant1979negation,loconte2025sum}. 

The constructions related to (decision) determinism in structured-decomposable circuits are closely connected to the recently published notion of partition circuits and partition trees~\cite{Zuidberg2024,martires2025quantum}. 
This line of work further motivates bridging quantum information theory with knowledge representation and compilation.

\section{Discussion}
\label{sec:discussion}

The established equivalences between tensor networks and tractable circuits provide a formal bridge between two communities that have developed largely in isolation: the tensor-network community (primarily in quantum physics, linear algebra, and machine learning) and the knowledge compilation community (primarily in formal verification, probabilistic reasoning, and automated reasoning).
The transformations between these two data structures with only linear (or quadratic) overhead enable direct transfer of e.g. lower bounds, heuristics, and analytical results in both directions. 

From a theoretical perspective, there is a large arsenal of tools for proving \emph{lower bounds} on circuits~\cite{Bova2016,deColnetM2020} based on rectangle cover techniques, rank- or width-based methods, etc. The underlying concepts of these tools are sometimes studied for tensor networks under different names, for instance the relationship between linear treewidth and Schmidt rank is well known~\cite{legeza2003optimizing}.
Our contributions suggest that these tools directly translate to the tensor network domain, as often used in other fields. For example, the DNNF lower bounds from~\cite{deColnet20} on permutation functions should also transfer to deterministic-TTN, which is useful since these functions correspond to Clifford operators, which are ubiquitous in, e.g., quantum error correction~\cite{cross2025small}.

There have been many heuristics for \emph{variable ordering optimization} for both circuits and MPS~\cite{legeza2003optimizing}, with strong evidence of \NP-hardness~\cite{bollig1996improving,hillar2013most}. For decision diagram circuits (BDD and SDD~\cite{Choi2013}), dynamic variable reordering has been extensively explored but remains largely unknown in the tensor network domain. Our mappings open possibilities for applying the techniques developed for decision diagrams to tensor networks.

The circuit community has spent efforts creating \emph{canonical tractable representations} of functions, in particular SDD and PSDD~\cite{BroeckD15}. Canonicity is a highly desirable property from a practical perspective. Canonical forms are also important for theoretical and numerical studies using tensor networks~\cite{Acuaviva2023}, but it is unknown how to compute them cheaply (SVD/Gaussian elimination is required~\cite{Schollwock_MPS_DMRG}). Our contribution changes this picture for deterministic-TTN. 
However, in practice there exist methods for structured problems to find optimal TTN~\cite{li2024optimal}, which is a weaker property than canonicity, which might be used for structured decomposable circuits.

There exist \emph{approximation procedures} for circuits and tensor networks  that could also be transferred from one domain to the other. 
On the tensor network side, a liberal form of approximation is often done through \emph{truncation}~\cite{Schollwock_MPS_DMRG}. Truncation has only recently been studied for decision diagrams~\cite{hillmich2022approximating}, but never for circuits before. On the circuit side, in Boolean settings, polytime methods for \emph{approximating \#P-hard problems} have been shown to apply to decomposable circuits~\cite{Meel2026,ArenasCJR21}. There is hope that these methods, known as FPRAS, are applicable to \#P-hard problems on real-valued decomposable circuits and, therefore, to TTN and MPS (tensor contraction, a key operation for tensor networks, is \#P-hard~\cite{biamonte2015tensor}).

The equivalences we found have strong implications for the \emph{succinctness and tractability} of these data structures and, therefore, when combined with the available knowledge compilation maps~\cite{DarwicheM02,fargier2014knowledge,vinkhuijzen2024knowledge}, can help choose an appropriate representation method for concrete cases, not only for circuits but also for tensor networks.

In summary, recent advances in knowledge compilation---efficiently computable canonical forms, lower bounds, parameterized complexity results, heuristics, and approximation schemes---can now benefit the tensor-network community, and vice versa. We expect that further exploration of these connections, including extensions to cyclic tensor networks such as PEPS or to unstructured circuits, will yield both theoretical insights and practical gains across machine learning, quantum computing, and automated reasoning.

\section*{AI Declaration}
In the preparation of this paper AI was only used to spot typos and improve readability.

\section*{Acknowledgments}
This publication is part of the project Divide \& Quantum  (with project number 1389.20.241) of the research programme NWA-ORC which is (partly) financed by the Dutch Research Council (NWO). This work was supported by the Netherlands Organization for Scientific Research (NWO/OCW), as part of QuantumLimits (project number SUMMIT.1.1016).

\bibliographystyle{kr}
\bibliography{main}

\end{document}